\documentclass[nato,numreferences]{crckbked}

\begin{document}

\begin{article}
\begin{opening}

\title{Classifying $N$-extended $1$-dimensional Supersymmetric Systems}

\author{Francesco \surname{Toppan}
\thanks{\email{toppan@cbpf.br}}}

\institute{CBPF, DCP, Rua Dr. Xavier Sigaud 150,\\ cep 22290-180
Rio de Janeiro (RJ), Brazil}

\runningauthor{Francesco Toppan} \runningtitle{Classifying
$N$-extended $1$-dimensional Supersymmetric Systems}

\end{opening}

\section{Introduction}

In this talk I will report some results obtained in a joint
collaboration with A. Pashnev, concerning the classification of
the irreducible representations of the $N$-extended Supersymmetry
in $1$ dimension and which find applications to the construction
of Supersymmetric Quantum Mechanical Systems \cite{pas/top}.
\par
This mathematical problem finds immediate application to the
theory of dimensionally (to one temporal dimension) supersymmetric
$4d$ theories, which gets $4$ times the number of supersymmetries
of the original models (the $N=8$ supergravity being e.g.
associated with the a $N=32$ Supersymmetric Quantum Mechanical
theory). Due to a lack of superfield formalism for $N>4$, only
partial results are known \cite{cro/rit} and \cite{cla/hal}.\par
More recently, Supersymmetric and Superconformal Quantum Mechanics
have been applied in describing e.g. the low-energy effective
dynamics of a certain class of black holes, for testing the
$AdS/CFT$ correspondence in the case of $AdS_2$, in investigating
the light-cone dynamics of supersymmetric theories.\par In this
report of the work with Pashnev, two main results will be
presented. At first a peculiar property of supersymmetry in one
dimension is exhibited, namely that any finite dimensional
multiplet containing $d$ bosons and $d$ fermions in different spin
states are put into classes of equivalence individuated by
irreducible multiplets of just two spin states, where all bosons
and all fermions are grouped in the same spin. Later it is shown
that all irreducible multiplets of this kind are in one-to-one
correspondence with the classification of real-valued Clifford
$\Gamma$ matrices of Weyl type.\par This classification refines
(in the case of ``non-Euclidean" supersymmetry, see below) the
results obtained in \cite{wit/tol/nic} and \cite{gat/ran}. Another
reference where some aspects of the theory of the representation
of $1$-dimensional supersymmetry are discussed is given by
\cite{col/pap}.\par The mathematical problem we are investigating
can be stated as follows, finding the irreducible representation
of the supersymmetry algebra
\begin{equation}\label{top-eq1}
\{ Q_i,Q_ j\}= {\omega}_{ij}H,
\end{equation}
where $Q_i,\;\; i=1,2,\cdots,N$ are supercharges
 and
\begin{equation}\label{top-eq2}
H= - i \frac{\partial }{ \partial t}
\end{equation}
is the Hamiltonian. The constant tensor  ${\omega}_{ij}$ can be
conveniently diagonalized and normalized in such a way to coincide
with a pseudo-Euclidean metric $\eta_{ij}$ with signature $(p,q).$
Usually the eigenvalues are all assumed being positive (i.e.
$q=0$), however examples can be given (see \cite{pas2}), of
physical systems whose supersymmetry algebra is characterized by
an indefinite tensor. In the following I will discuss the simplest
example of this kind.\par Any given finite-dimensional
representation multiplet of the above superalgebra can be
represented in form of a chain of $d$ bosons and $d$ fermions
\begin{equation}\label{top-eq3}
\Phi^0_{a_0}, \quad \Phi^1_{a_1}, \quad \cdots ,\quad
\Phi^{M-1}_{a_{M-1}}, \quad \Phi^M_{a_M}
\end{equation}
whose components  $\Phi^I_{a_I}$, $ (a_I=1,2,\cdots,d_I)$ are real
and alternatively bosonic and fermionic
($d=d_0+d_2+d_4+...=d_1+d_3+d_5+...$). For such a multiplet the
short notation ${\bf \{d_0, d_1, \cdots ,d_{M}\}}$ will also be
employed.\par Due to dimensionality argument the $i-th$
supersymmetry transformation for the ${\Phi^I}_{a_I}$ components
is given by
\begin{equation}\label{top-eq4}
  \delta_\varepsilon \Phi^I_{a_I} = \varepsilon^i
{(C^I_i)_{a_I}}^{a_{I+1}}\Phi^{I+1}_{a_{I+1}} + \varepsilon^i
{(\tilde{C}^I_i)_{a_I}}^{a_{I-1}}\frac{d}{d\tau}{\Phi}^{I-1}_{a_{I-1}},
\end{equation}
and it simplifies for the end-components (due to the absence of
the $I=-1$ and $I=M+1$ components).\par In one dimension it is
therefore possible to redefine the last components according to
\begin{equation}\label{top-eq5}
  \Phi^M_{a_M}=\frac{d}{d\tau}\Psi^{M-2}_{a_M}
\end{equation}
in terms of some functions $\Psi^{M-2}_{a_M}$. The initial
supermultiplet of length $M+1$ is now re-expressed as the ${\bf
\{d_0, d_1, \cdots, d_{M-2}+d_M, d_{M-1}, 0\}}$ supermultiplet of
length $M$. By repeating $M$ times the same procedure the shortest
supermultiplet ${\bf \{d,d\}}$ of length $2$ can be reached. The
above argument outlines the proof of the statement that all
supermultiplets are classified according to the irreducible
representations of supermultiplets of length $2$.

\section{Extended supersymmetries and real valued Clifford
algebras}

The main result of the previous Section is that the problem of
classifying all $N$-extended supersymmetric quantum mechanical
systems is reduced to
 the problem of classifying the irreducible
representations of length $2$. Having this in mind let us simplify
the notations. Let the indices
 $a, \alpha = 1,
\cdots, d$ number the bosonic (and respectively fermionic)
elements in the SUSY multiplet. All  of them are assumed to depend
on the time coordinate $\tau$ ($X_a\equiv X_a(\tau)$,
$\theta_\alpha\equiv \theta_\alpha (\tau)$).
\par
In order to be definite and without loss of generality let us take
the bosonic elements to be the first ones in the chain ${\bf
\{d,d\}}$, which can be conveniently represented also as a column
\begin{equation}\label{top-eq6}
\Psi=\left(
\begin{array}{c}
X_a \\ \theta_\alpha
\end{array}
\right),
\end{equation}
the supersymmetry transformations are reduced to the following set
of equations
\begin{eqnarray}\label{top-eq7}
\delta_\varepsilon X_a &=& \varepsilon^i
{(C_i)_a}^{\alpha}\theta_{\alpha}\equiv i(\varepsilon^iQ_i \Psi)_a
\nonumber\\ \delta_\varepsilon \theta_{\alpha} &=&\varepsilon^i
{(\tilde{C}_i)_{\alpha}}^b\frac{d}{d\tau} X_b\equiv
i(\varepsilon^iQ_i \Psi)_\alpha
\end{eqnarray}
where, as a consequence of (\ref{top-eq1}),
\begin{eqnarray}\label{top-eq8}
C_i {\tilde C}_j + C_j{\tilde C}_i &=& i \eta_{ij}
\end{eqnarray}
and
\begin{eqnarray}\label{top-eq9}
{\tilde C}_i C_j + {\tilde C}_j C_i &=& i\eta_{ij}
\end{eqnarray}
Since $\varepsilon_i, X_a, \theta_\alpha$ are  real, the matrices
 $C_i$'s, ${\tilde C}_i$'s have to be respectively
imaginary and real.
 If we set (just for normalization)
\begin{eqnarray}
C_i &=&  \frac{i}{\sqrt{2}} \sigma_i\nonumber\\\label{top-eq10}
 {\tilde C}_i&=&\frac{1}{\sqrt{2}}{\tilde\sigma}_i
\end{eqnarray}
and accommodate $ \sigma_i, {\tilde\sigma}_i$ into a single matrix
\begin{equation}\label{top-eq11}
\Gamma_i=\left(
\begin{array}{cc}
0 & \sigma_i \\ {\tilde\sigma}_i& 0
\end{array}
\right),
\end{equation}
they form a set of real-valued Clifford $\Gamma$-matrices of Weyl
type (i.e. block antidiagonal), obeying the (pseudo-) Euclidean
anticommutation relations
\begin{eqnarray}\label{top-eq12}
\{\Gamma_i, \Gamma_j \} &=& 2 \eta_{ij}.
\end{eqnarray}
Therefore the classification of irreducible multiplets of
representation of a $(p,q)$ extended supersymmetry is in
one-to-one correspondence with the classification of the
real-valued Clifford algebras $C_{p,q}$ with the further property
that the $\Gamma$ matrices can be realized in Weyl (i.e. block
antidiagonal) form.\par Real-valued Clifford algebras have been
classified in \cite{ati/bot/sha} for compact ($q=0$) case, and in
\cite{por} for the non-compact one. I follow here the exposition
in \cite{oku}. \par Three cases have to be distinguished for real
representations, specified by the type of most general solution
allowed for a real matrix $S$ commuting with all the Clifford
$\Gamma_i$ matrices, i.e.\\
 {\em i)} the normal case, realized
when $S$ is a multiple of the identity, \\ {\em ii)} the almost
complex case, for $S$ being given by a linear combination of the
identity and of a real $J^2= -{\bf 1}$ matrix,\\ {\em iii)}
finally the quaternionic case, for $S$ being a linear combination
of real matrices satisfying the quaternionic algebra.
\par
Real irreducible representations of normal type exist whenever the
condition\\ $p-q = 0,1,2 \quad mod \quad 8$ is satisfied (their
dimensionality being given by $2^{[\frac{N}{2}]}$, where $N=p+q$),
while the almost complex and the quaternionic type representations
 are realized in the $p-q
= 3,7 \quad mod \quad 8$ and in the $p-q = 4,5,6 \quad mod \quad
8$ cases respectively. The dimensionality of these representations
is given in both cases by $2^{[\frac{N}{2}]+1}$.\par We further
require the extra-condition that the real representations should
admit a block antidiagonal realization for the Clifford $\Gamma$
matrices. This condition is met for $p-q = 0 \quad mod \quad 8$ in
the normal case (it corresponds to the standard Majorana-Weyl
requirement), $p-q = 7 \quad mod \quad 8$ in the almost complex
case and  $p-q = 4,6 \quad mod \quad 8$ in the quaternionic case.
In all these cases the real irreducible representation is unique.
\par
It is therefore possible to furnish the dimensionality of the
irreducible representations of the of the supersymmetry algebra
or, conversely, the allowed $(p,q)$ signatures associated to a
given dimensionality of the bosonic and fermionic spaces. The
latter result is conveniently expressed by introducing the notion
of maximally extended supersymmetry. The $C_{p,q}$ ($p-q= 6\quad
mod \quad 8$) real representation for the quaternionic case can be
recovered from the $7 \quad mod \quad 8$ almost complex
$C_{p+1,q}$ representation by deleting one of the $\Gamma$
matrices; in its turn the latter representation is recovered from
the $C_{p+2,q}$ normal Majorana-Weyl representation by deleting
another $\Gamma$ matrix. The dimensionality of the three
representations above being the same, the normal Majorana-Weyl
representation realizes the maximal possible extension of
supersymmetry compatible with the dimensionality of the
representation. In search for the maximal extension of
supersymmetry we can therefore limit ourselves to consider the
normal Majorana-Weyl representations, as well as the quaternionic
ones satisfying the $p-q = 4\quad mod \quad 8$ condition.\par Let
us therefore introduce a parameter $\epsilon$, which assumes two
values and is used to distinguish the Majorana-Weyl ($\epsilon =0
$) with respect to the quaternionic case ($\epsilon = 1$). A space
of $d=2^t$ bosonic and $d= 2^t$ fermionic states can carry the
following set of maximally extended supersymmetries
\begin{equation}\label{top-eq13}
(p= t-4z + 5 - 3\epsilon, q= t+4z+\epsilon - 3)
\end{equation}
where the integer $z= k-l$ must take values in the interval
\begin{equation}\label{top-eq14}
\frac{1}{4}(3-t-\epsilon ) \leq z \leq \frac{1}{4} (t+ 5-
3\epsilon )
\end{equation} in order to guarantee the $p\geq 0$
and $q\geq 0 $ requirements.

\section{An application and conclusions.}

One of the most significant application of extended supersymmetric
quantum mechanics concerns the $1$-dimensional $\sigma$ models
evolving in a target spacetime manifold presenting both bosonic
and fermionic coordinates. In general such models present a
non-linear kinetic term and the extended supersymmetries put
constraints on the metric of the target. In this section let us
present here a very simplified model, which however is
illustrative of how invariances under pseudo-Euclidean
supersymmetry can arise. Let us in fact consider a model of $d$
bosonic fields $X_a$ and $d$ spinors $\psi_\alpha$ freely moving
in a flat $d$-dimensional target manifold, not necessarily
Minkowskian or Euclidean, endorsed of a pseudo-euclidean
$\eta_{ab}$. Let us furthermore introduce the free kinetic action
being given by
\begin{eqnarray}\label{top-eq15}
S_K=\int dt{\cal L} &=& \frac{1}{2}\int dt \left( \dot{X}_a
\dot{X}_b \eta^{ab} +i\delta
\dot{\psi}_\alpha\psi_\beta\eta^{\alpha\beta}\right),
\end{eqnarray}
where the metric $\eta^{\alpha\beta}$ for the spinorial part is
assumed to have the same signature as the metric $\eta^{ab}$, and
$\delta$ is just a sign normalization ($\delta=\pm 1$).
\par
A natural question to be asked is which supersymmetries are
invariances of the above free kinetic action. The answer is
furnished by accommodating the $d$ bosonic and $d$ fermionic
coordinates into a (maximally extended) irreducible representation
of the extended supersymmetries, and later counting how many such
transformations survive as invariances of the action. The first
non-trivial example concerns a $2$-dimensional target($d=2$),
whose two bosonic and two fermionic degrees of freedom carry the
${\bf \{2,2\}}$ representation of $(2,2)$ extended supersymmetry.
However, only half of these supersymmetries are realized as
invariances of the action. The action indeed is invariant under
either the $(2,0)$ or the $(1,1)$ extended supersymmetries,
whether the target space is respectively Euclidean or Minkowskian.
Therefore already in the $2$-dimensional Minkowskian case we
observe the arising of a pseudo-Euclidean supersymmetry
invariance. The next simplest example is realized by a
$4$-dimensional target. The four bosonic and four fermionic
coordinates can be accommodated into three irreducible
representations of maximally extended supersymmetry, according to
formula (\ref{top-eq13}), namely the $(4,0)$, the $(0,4)$ and the
$(3,3)$ extended supersymmetries. The action (\ref{top-eq15})
turns out to be invariant, for Euclidean $(4+0)$, Minkowskian
$(3+1)$ and $(2+2)$ signature for the metric $\eta$, according to
the following table
\begin{center}
\begin{tabular}{|c|c|c|c|c|}
 \hline  & $(4,0)$ & $(0,4)$ & $(3,3)$ &  \\
  \hline $(4+0)$ & $(4,0)$ & $(0,0)$ & $(3,0)$ & $\delta=+1$ \\
   $(4+0)$ & $(0,0)$ & $(0,4)$ & $(0,3)$ & $\delta =-1 $\\
  \hline $(3+1)$ & $(1,0)$ & $(0,0)$& $(1,0)$ & $\delta=+1$ \\
   $(3+1)$ & $(0,0)$ & $(0,1)$ & $(0,1)$ & $\delta=-1$ \\
  \hline $(2+2)$ & $(2,0)$ & $(0,2)$ & $(2,1)$ & $\delta=+1$\\
  $(2+2)$ & $(2,0$& $(0,2)$& $(1,2)$& $\delta=-1$\\\hline
\end{tabular}
\end{center}
which should be understood as follows. The central entries denote
how many supersymmetries are realized as invariances of the
(\ref{top-eq15}) action for each one of the three irreducible
representations of maximally extended supersymetry, in
correspondence with the given signature of spacetime and sign for
$\delta$. In this particular case invariance under
pseudo-Euclidean supersymmetry is guaranteed for the target of
signature $(2+2)$.\par In this talk I have presented some results
concerning the representation theory for irreducible multiplets of
the one-dimensional $N=(p,q)$ extended supersymmetry. A peculiar
feature of the one-dimensional supersymmetric algebras consists in
the fact that the supermultiplets formed by $d$ bosonic and $d$
fermionic degrees of freedom accommodated in a chain with $M+1$
$(M\geq 2)$ different spin states uniquely determines a $2$-chain
multiplet of the form ${\bf \{d,d\}}$ which carries a
representation of the $N$ extended supersymmetry. Furthermore, it
is shown that all such $2$-chain irreducible multiplets of the
$(p,q)$ extended supersymmetry are fully classified; when e.g. the
condition $p-q=0$ mod $8$ is satisfied, their classification is
equivalent to that one of Majorana-Weyl spinors in any given
space-time, the number $p+q$ of extended supersymmetries being
associated to the dimensionality $D$ of the spacetime, while the
$2d$ supermultiplet dimensionality is  the dimensionality of the
corresponding $\Gamma$ matrices. The more general case for
arbitrary values of $p$ and $q$ has also been fully discussed.
\par
These mathematical properties can find a lot of interesting
applications in connection with the construction of Supersymmetric
and Superconformal Quantum Mechanical Models. These theories are
vastly studied due to their relevance in many different physical
domains, to name just a few it can be mentioned the low-energy
effective dynamics of black-hole models, the dimensional reduction
of higher-dimensional superfield theories, which are a laboratory
for the investigation of the  spontaneous breaking of the
supersymmetry, and so on. \vspace{.5cm}\\ \noindent{{\bf
Acknowledgments.}} It is a pleasure for me to acknowledge A.
Pashnev. The results reported in this talk are fruit of our
collaboration. I wish also acknowledge for useful discussions E.A.
Ivanov, S. J. Gates Jr., S.O. Krivonos and V. Zima. Finally, let
me express my gratitude to the organizers of the ARW conference
for the invitation and the warm hospitality.

\end{article}


\begin{thebibliography}{1}
\bibitem{pas/top}
A. Pashnev and F. Toppan, {\it On the Classification of
$N$-Extended Supersymmetric Quantum Mechanical Systems}, {\sl
CBPF, JINR} {\it preprint}, CBPF-NF-029/00, JINR E2-2000-193, {\tt
hep-th/0010135}, Dubna, Rio de Janeiro, 2000.
\bibitem{cro/rit} M. De Crombrugghe and V. Rittenberg,
\newblock Ann. of Phys. \newblock {\bf 151} (1983), 99.
\bibitem{cla/hal} M.Claudson and M.B. Halpern,
\newblock Nucl.Phys.
\newblock {\bf B250} (1985), 689.
\bibitem{wit/tol/nic} B. de Wit, A.K. Tollsten and H. Nicolai,
\newblock
Nucl.Phys. \newblock {\bf B392} (1993), 3.
\bibitem{gat/ran} S. James Gates, Jr. and Lubna Rana,
\newblock Phys. Lett.
\newblock {\bf B352} (1995), 50;
\newblock ibid.
\newblock {\bf B369} (1996), 262.
\bibitem{col/pap}
R.A. Coles and G. Papadopoulos,
\newblock Class. Quant. Grav.
\newblock {\bf 7} (1990), 427--438.
\bibitem{pas2}
A. Pashnev, {\it Noncompact Extension of One-Dimensional
Supersymmetry and Spinning Particle}, {\sl JINR} {\it preprint},
E2-91-536, Dubna, 1991.
\bibitem{ati/bot/sha} M. Atiyah, R. Bott and A. Shapiro,
\newblock Topology
\newblock{\bf 3}, (Suppl. 1) (1964), 3.
\bibitem{por} I. Porteous,
\newblock{\it Topological Geometry},
\newblock van Nostand Rheinhold, London, (1969).
\bibitem{oku} S. Okubo,
\newblock Jou. Math. Phys., \newblock
{\bf 32} (1991), 1657; ibid. 1669.
\end{thebibliography}
\end{document}